\documentclass[aps,prb,twocolumn,superscriptaddress,longbibliography, amsmath,amssymb,amsfonts,citeautoscript]{revtex4-2}
\usepackage{graphicx}
\usepackage{epsf}
\usepackage{epstopdf}
\usepackage{graphicx}
\usepackage{tikz}
\usepackage{amsmath,bm,upgreek}
\usepackage{nccmath}
\usepackage[mathscr]{euscript}
\usepackage{natbib}
\usepackage{hyperref}

\usepackage{geometry}
\usetikzlibrary{positioning, arrows.meta}

\usepackage[utf8]{inputenc}
\usepackage{amsmath}
\usepackage{amsfonts}
\usepackage{amssymb}
\usepackage{geometry}
\usepackage{hyperref}
\usepackage{booktabs}

\usepackage{amsmath,amssymb,amsfonts}
\usepackage{bm}
\usepackage{physics}
\usepackage{hyperref}
\usepackage[normalem]{ulem}

\graphicspath{{./figs/}}
\begin{document}

\title{Search for Majorana Bound States in Short Chains of Proxmitised Quantum Dots}

\author{Bogdan R. Bu{\l}ka}
\email{bulka@ifmpan.poznan.pl}
\affiliation{Institute of Molecular Physics, Polish Academy of
Sciences, ul. M. Smoluchowskiego 17, 60-179 Pozna{\'n}, Poland
}
\author{Karol I. Wysoki\'nski}
\email{karol.wysokinski@mail.umcs.pl}

\affiliation{Institute of Physics, M. Curie-Sk\l{}odowska University, 20-031 Lublin, Poland}

\begin{abstract}

Majorana zero modes (MZM) appearing at the ends of artificially created one-dimensional p-wave superconductors have been intensively studied recently both theoretically and experimentally. Among possible platforms, proximitised semiconducting wires, and short chains of quantum dots with a superconductor in between were investigated. Here, we propose a different platform consisting of a chain of quantum dots (QDs) sandwiched between an s-wave superconductor and a strong spin-orbit semiconductor, subject to a Zeeman magnetic field.
Neglecting spin-conserving hopping processes between QDs and local on-dot superconducting correlations induced by the superconducting proximity effect,  reduces the Hamiltonian to the sum of two equivalent Hamiltonians with two independent Hilbert spaces.
The resulting model has a staggered structure due to spin-flipping processes $t_{so}$ and cross-Andreev reflections $\Delta_{CAR}$ between neighbouring dots.
Our central result is the phase diagram of a short chain consisting of four QDs and coupled to two external reservoirs, obtained by means of the Green function in chiral Majorana representation.
The modulus of the retarded Green function, probing the whole chain and calculated for zero energy, is shown to contain information on topology and spatial character of Majorana zero modes. The features observed in the Green function nicely agree with those obtained from the transfer matrix approach. In particular the region in parameter space in which Majorana zero modes display  oscillatory wave functions are well reproduced.
Likewise the borders of the fermion parity changes obtained by the Green function agree with those obtained by other means.

\end{abstract}

\maketitle

\section{Introduction}

 Majorana zero modes have been proposed as building blocks of topological quantum computers~\cite{Kitaev2001,Alicea2012}. They are expected to exist at the edges of topological superconductors or at the ends of topological superconducting wires~\cite{Lutchyn2010,Oreg2010}. Theoretically, their existence and topological protection have been established for the Kitaev chain of spinless electrons~\cite{Bernevig2013}. The  Kitaev chain realised by using a semiconducting nano-wire with strong spin-orbit interaction, placed in a magnetic field and contacted to an s-wave superconductor provides ambiguous results due to unavoidable disorder leading to the appearance of spurious zero modes~\cite{Pan2021}.

These difficulties have resulted in a proposal to replace a semiconducting wire with two quantum dots coupled to a superconductor and in this way realise a two-site Kitaev chain~\cite{Sau2012,Leijnse2012}. Such chain contains zero-energy end-modes, which are known as {\it poor man Majorana} (PMM) quasiparticles, because they are not protected and exist only at a special point in parameter space. Recently, such a minimal Kitaev chain has been realised~\cite{Dvir2023}, opening the door to novel platforms with a large degree of control.
The experimental modifications of the original set-up of Dvir {\it et al.}~\cite{Dvir2023} have been proposed~\cite{Tsintzis2022}, and the extension to three-site long chains realised~\cite{Bordin2024,Bordin2025}. At present, however, it is not clear if longer chains will lead to more stable Majorana quasiparticles due to odd-even effects~\cite{Ezawa2024}, longer range hopping, unavoidable inhomogeneities ~\cite{Svensson2024,Miles2024} or strong interaction which may induce zero modes of slightly different character~\cite{Bozkurt2025}.
On the other hand, non-hermiticity has been argued to stabilise zero-energy Majorana modes~\cite{Cayao2025}. Non-hermiticity naturally appears in short Kitaev chains due to the strong coupling of a chain to normal reservoirs. Technically, it is induced by the effective self-energy terms due to the dots-leads couplings. Such terms naturally exist in all transport measurements. In the weak coupling, the role of leads is limited to smearing of various features in the appropriate Green function.

In most theoretical proposals to realise minimal Kitaev chains with the help of QDs~\cite{Leijnse2012,Aasan2016,Luethi2024,Luethi2025}, two external dots are coupled to a third central one, which is contacted to a superconductor; eventually, quantum dots are coupled to a common superconductor placed between them. Two and three-dot structures of a similar type were recently realised experimentally, and the Majorana modes were found under appropriate conditions~\cite{Bordin2024,Bordin2025,Bordoloi2022,Zatelli2024}.

In a previous paper~\cite{Bulka2026}, we proposed a minimal two-dot Kitaev chain in which both quantum dots are sandwiched between an \textit{s}-wave superconductor and a semiconductor with strong spin--orbit interaction. In this setup, two dual processes are essential: crossed Andreev reflection (CAR) and spin--orbit interaction (SOI). Their interplay leads to the formation of fully spin-polarized zero-energy Majorana bound states localized on different quantum dots.

In the present work, we extend this concept to chains containing many quantum dots arranged in the same sandwich geometry.
Our objective is to check if and how the topological properties develop with increasing chain length.
By formulating the model for chains with an even number of dots we observe its staggered character, allowing the introduction of two sublattices and a corresponding chiral Majorana representation.
This formulation provides a transparent framework for comparing the shortest nontrivial chain with the infinite system and for identifying the finite-size precursors of the bulk topological phase. In particular, it enables us to obtain information about the degree of topology in short chains.
As the simplest finite example, we consider a four-dot structure whose end dots are coupled to two external metallic leads. The leads act as dissipative environments and allow the injection and extraction of charge from the system. They also introduce dissipation into the system, which turns out to be convenient for calculating retarded Green functions.

The organisation of the rest of the paper is as follows. In Section \ref{sec:n-QDs} we introduce our model, calculate the spectrum of a chain with $n=2,4,6$ dots and consider protection of the zero-energy mode.  Section \ref{sec:majoranarepr} introduces Majorana representation of the $n=4$ chain coupled to external electrodes.   Topological aspects of the short chain are discussed in Section \ref{sec:topo-phase} by comparing the features of the Majorana Green function of the short chain ($n=4$)  with those obtained by the transfer matrix approach, including spatial oscillatory solutions and level parity inversion, as well as the expected phase diagram of the long chain $n\rightarrow \infty$. Some details are relegated to three Appendices. We sum up and conclude in Section \ref{sec:concl}.

\section{Proximitised chain of n QDs: the Hamiltonian and  spectrum }\label{sec:n-QDs}

\begin{figure}
\centerline{\includegraphics[width=0.7\linewidth,clip]{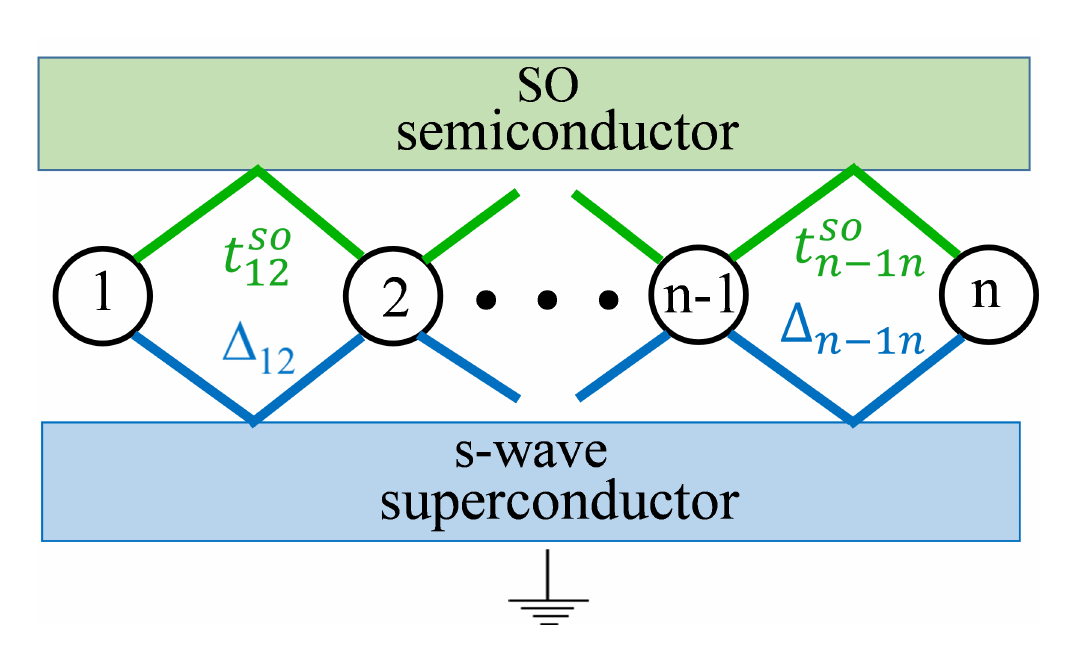}}
\caption{Scheme of a sandwich geometry of n quantum dots chain in proximity to the BCS superconductor and a semiconductor with strong spin - orbit coupling.  }
\label{scheme1}
\end{figure}

We start with the system shown in Fig. \ref{scheme1} consisting of a standard fully gapped s-wave superconductor (with a gap of magnitude $\Delta_S$) and a chain of quantum dots in proximity to it, and a semiconductor with strong spin-orbit coupling. Each of the bulk quantum dots is tunnel-coupled to two neighbouring dots.

As we are interested in low-energy processes only, we shall assume a large superconducting gap limit. It means we consider proximitised quantum chain in which each quantum dot inherits superconducting correlations due to its coupling to the superconductor. The induced superconducting correlations result from two kinds of processes between the superconductor and quantum dots, known as Andreev reflections. First, the direct Andreev reflection processes. In the following, we call them local Andreev reflections (LAR), and they induce on-dot pairing with $\Delta_{LAR}$. The crossed Andreev reflections involve two neighbouring QDs and shall be denoted as CAR-processes. They induce interdot pairing correlations with the amplitude $\Delta_{CAR}$.
Importantly, these cross Andreev reflections may be effective not only between neighbouring quantum dots. However, the dots involved in the process have to be not very far apart; at a distance smaller than the superconducting coherence length $\xi$. On the one side this requirement introduces some limitations for the model; on the other it may be used as an additional means to realise the desired structure.

The effective Hamiltonian describing the proximitised quantum dot chain is similar to that studied earlier~\cite{Bulka2026}, here written for an arbitrary number of quantum dots

\begin{align}
H_{prox} &=\sum_{i\sigma} (\epsilon_i-\mu) n_{i\sigma} +V_z\sum_{i}(d^\dagger_{i\uparrow}d_{i\uparrow}-d^\dagger_{i\downarrow}d_{i\downarrow}) \nonumber\\
&-\Delta_{LAR} \sum_{i} (d^{\dag}_{i\uparrow}d^{\dag}_{i\downarrow}+d_{i\downarrow}d_{i\uparrow})\nonumber\\
&-\Delta_{CAR}\sum_{i}(d^{\dag}_{i\uparrow}d^{\dag}_{i+1\downarrow}-d^{\dag}_{i\downarrow}d^{\dag}_{i+1\uparrow} +\text{H.c.})\nonumber\\
&+t\sum_{i\sigma} d^\dagger_{i\sigma}d_{i+1\sigma}\nonumber \\
&+\sum_{i}t_{so}(d^\dagger_{i\uparrow}d_{i+1\downarrow}
-d^\dagger_{i\downarrow}d_{i+1\uparrow}+ \text{H.c.}).
\label{ham:prox}
\end{align}
The operator  $d^\dagger_{i\sigma}$, creates the spin $\sigma$
electron on dot $i$, while the operator $d_{i\sigma}$ denotes the corresponding annihilation operator. In the spirit of tight-binding approximation, we assumed both hopping $t$ and spin-orbit hopping $t_{so}$.

The Hamiltonian (\ref{ham:prox}) serves as a starting point of our analysis. It is an effective Hamiltonian of the original sandwich shown schematically in Fig. \ref{scheme1} with the superconductor's degrees of freedom integrated out. The meaning of symbols is standard. The dots' energy levels are denoted by $\epsilon_i$ and the chemical potential by $\mu$. $V_z$ denotes the Zeeman splitting, which we assume to be constant in magnitude along the chain, local and non-local pairing correlations are unavoidable in our geometry, with all quantum dots in close proximity to the superconductor.

Two different normal hopping processes in the above Hamiltonian are described by parameters $t$ and $t_{so}$. The former describes standard hopping in which electron transits between the $i$-th and $i+1$-th dot, while the latter describes hopping with the rotation of spin by $\pi$.

Considering two Hilbert spaces ($c.f.$  \ref{sec:twoHilbert}) one observes that it is the terms $t$ and $\Delta_{LAR}$ which mix them.
Neglecting both terms allows us to write the Hamiltonian (\ref{ham:prox}) as a sum of two Hamiltonians, and each of which describes spin staggered chains
\begin{align}
H_{\text{prox}} = H_{2n\Psi} + H_{2n\Phi}.
\end{align}
Defining a single site Nambu spinor  $c^{\dag}_{i\sigma} = (d^{\dag}_{i\sigma},\, d_{i\sigma})$ allows to write
\begin{align}
H_{2n\Psi} &= \sum_{j=0}^{2n-1}
\Big[
(c^{\dag}_{2j+1\uparrow}\check{Q}c_{2j+2\downarrow}
-
c^{\dag}_{2j+2\downarrow}\check{Q}c_{2j+3\uparrow})
+ \text{h.c.}
\Big]
\nonumber\\
&+\frac{1}{2}\sum_{j=0}^{2n-1}
\Big[
\epsilon_{\uparrow}c^{\dag}_{2j+1\uparrow}\tau_{z} c_{2j+1\uparrow}
+
\epsilon_{\downarrow}c^{\dag}_{2j+2\downarrow}\tau_{z} c_{2j+2\downarrow}
\Big],
\label{hN2}
\end{align}
This Hamiltonian is conveniently rewritten in the Bogoliubov-de Gennes (BdG) form as
\begin{equation}
H_{2n\Psi}
=
\frac{1}{2}
\Psi^\dagger
\mathcal H_{2n\Psi}
\Psi,
\end{equation}
where the Nambu spinor of the whole chain is defined in terms of the single site spinors as
$\Psi^T =
\begin{pmatrix}
c_{1\uparrow},\,
c_{2\downarrow} ,\,
c_{3\uparrow} ,\,
c_{4\downarrow},\,
\cdots
\end{pmatrix}
$
and
\begin{equation}
\mathcal H_{2n\Psi} =
\begin{pmatrix}
\epsilon_{\uparrow}\tau_z & \check Q & 0 &0 &0&\cdots \\
\check Q^T & \epsilon_{\downarrow}\tau_z & -\check Q & 0 &0&\cdots\\
0 & -\check Q^T & \epsilon_{\uparrow}\tau_z & \check Q &0&\cdots\\
0 & 0 & \check Q^T & \epsilon_{\downarrow}\tau_z &  -\check Q&\cdots\\
\vdots&\vdots&\vdots&\vdots&\vdots&\ddots
\end{pmatrix}.
\label{eq:HNPsi}
\end{equation}
In the following, we investigate only the Hamiltonian $\mathcal{H}_{2n\Psi}$, because the properties of $\mathcal{H}_{2n\Phi}$ can be deduced by reversing the spin orientation, and adequately changing the corresponding model parameters. The model (\ref{eq:HNPsi})
describes a staggered chain with the alternating hopping $ \check Q = t_{so}\tau_z - \Delta_{CAR}(i\tau_y)$ and the staggered onsite potential $\epsilon_{\uparrow}=m_A=\mu+V_z$, $\epsilon_{\downarrow}=m_B=\mu-V_z$.
This feature calls for a description of the model using two sublattices, A and B. In this description, both spin-flipping hopping between consecutive dots $t_{so}$ and the anomalous hopping $\Delta_{CAR}$ contribute to the transfer of electrons along the chain.

\begin{figure}
\includegraphics[width=0.8\linewidth,clip]{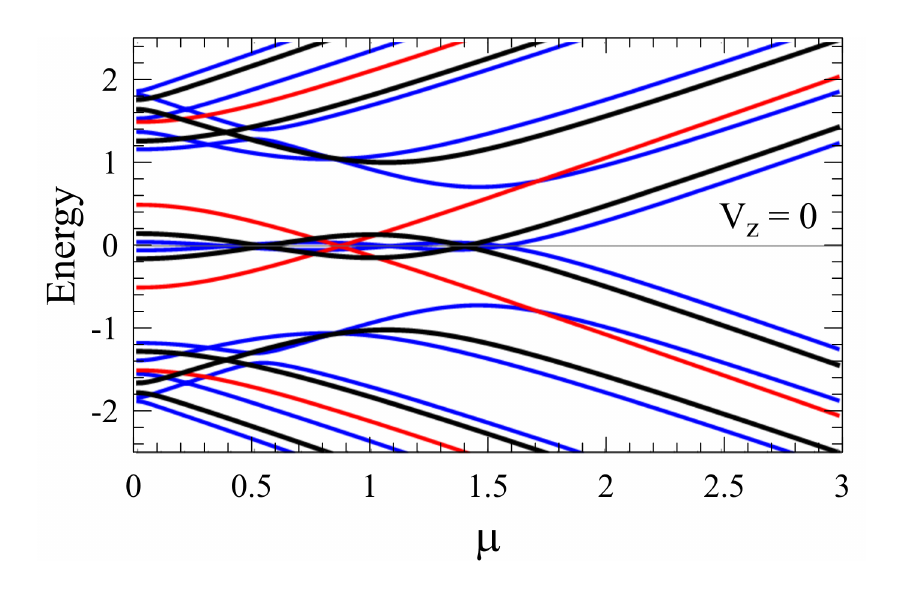}
\includegraphics[width=0.8\linewidth,clip]{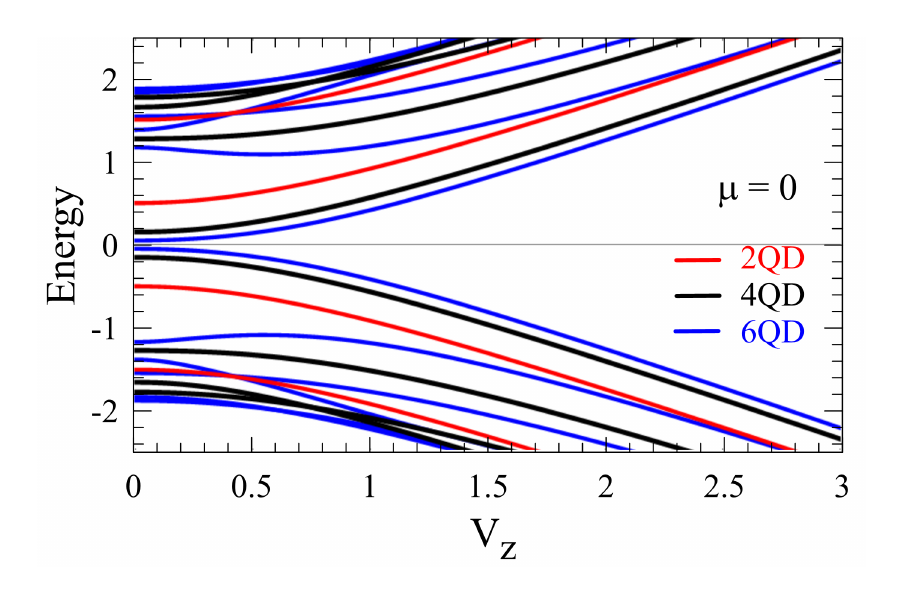}
\caption{Energy spectrum of the chain with 2QD (red),  4QD (black), and 6QD (blue) as a function of $\mu$ at $V_z=0$ (top panel) and a function of $V_z$ at $\mu=0$  (bottom panel) for  $t_{so}=1$ and $\Delta_{CAR}=0.5$.  Note the level crossing near zero energy, as well as the crossings in the upper and lower energy branches.
\label{spectrum}}
\end{figure}

The spectrum of the finite chain consisting of 2n-QDs can be calculated from  the zeros of the determinant
\begin{align}
d_{2nQD} \equiv \;\det[\omega-{H}_{2n\Psi}]=0.
\end{align}
We have calculated the spectra for 2QD, 4QD and 6QD as a function of the chemical potential, which in real systems can be changed by the external gates attached to DQs and as a function of the amplitude of the Zeeman field $V_z$. The spectra are shown in  Fig.~\ref{spectrum}. They are obtained for $\Delta_{CAR}=0.5$ and $t_{so}=1$. The left panel shows the changes in the spectrum obtained for $V_z=0$ with increasing $\mu$. One observes a pair of states with energies oscillating close to $E=0$. They cross each other at $E=0$ with increasing $\mu$. Importantly, the amplitude of oscillations diminishes with the number $n$ of QDs in a chain. Also, the number of crossings equals the number of dimmers in a chain, and thus, for a chain with six dots (three dimers), they cross three times. These features are signatures of the zero-energy Majorana state appearing in a long chain with $n\rightarrow\infty$. The magnitude of splitting of the lowest (zero) energy states depends on the overlap between zero-energy modes at the chain ends. For longer chains, the overlap gets smaller, and the splitting decreases. For $\mu$ exceeding some ``critical'' value these low-energy states move apart, and a robust energy gap appears. The ``critical'' values of the chemical potential above which the gaps in the spectrum appear depend on $n$ and grow with increasing $n$. This dependence is rather weak for large $n$.

The right panel of Fig.~\ref{spectrum} illustrates the effect of the Zeeman field on the spectrum calculated for $\mu=0$. We observe the monotonic increase of the minigap from its initial value.
The lowest energy states monotonously depart with increasing $V_z$ from the values they have for a given $n$ and $\mu=0$. The dependence of the higher energy states on $V_z$ is more complicated, with a few level crossings clearly visible for $n=6$ chain.

The sweet spot of the model corresponds~\cite{Bulka2026} to $t_{so}=\Delta_{CAR}=1$. For those values of parameters, the lowest eigen-energies calculated for $\mu=0=V_z$ equal zero. The dependence of the spectrum on $\mu$ or $V_z$ is similar to that shown in Fig.~\ref{spectrum}, except that all states evolve from zero in a monotonic manner. The critical values of the chemical potential and Zeeman field increase with the chain length $n$.

As our interest is in the zero-energy Majorana bound state and its protection, we calculated the corresponding eigenvalues $\omega^0_{nQD}$ and expanded them in series with respect to the model parameters \cite{Leijnse2012}.
The robustness with respect to the perturbation $V_z>0$ and the deviation from the sweet spot $t_{so}-\Delta_{CAR}$ depends on the length of the chain.
The results for 2QD, 3QD and 4QD chains read
\begin{align}
|\omega_{2QD}^0|\propto\; &
(t_{so}-\Delta_{CAR})+\frac{V_z^2}{2t_{so}}\nonumber\\
&+\mathcal{O}[\frac{V_z^3}{t_{so}^3}, \frac{(t_{so}-\Delta_{CAR})^2}{t_{so}^2}],\\
|\omega_{3QD}^0|\propto\; & \frac{(t_{so}-\Delta_{CAR})V_z}{t_{so}}+
\frac{V_z^3}{4t_{so}^2}\nonumber\\
&+\mathcal{O}[\frac{V_z^4}{t_{so}^4},\frac{(t_{so}-\Delta_{CAR})^2}{t_{so}^2}],\\
|\omega_{4QD}^0|\propto\;  & \frac{(t_{so}-\Delta_{CAR})^2}{2t_{so}}+\frac{3(t_{so}-\Delta_{CAR})V_z^2}{4t_{so}^2}\nonumber\\
&+
\frac{V_z^4}{8t_{so}^3}+\mathcal{O}[\frac{V_z^5}{t_{so}^5},\frac{(t_{so}-\Delta_{CAR})^3}{t_{so}^3}].
\end{align}
Notice that at the sweet spot, the protection of the zero-energy state increases as $|\omega_{nQD}^0|\propto\ V_z^n/(2t_{so})^{n-1}$ with the number $n$ of QDs. The protection also increases with the deviation from the sweet spot $t_{so}=\Delta_{CAR}$ as  $|\omega_{nQD}^0|\propto\; (t_{so}-\Delta_{CAR})^{n/2}/(2t_{so})^{n/2-1}$ for the even number $n$, while for the odd number $n$ the zero-energy state is trivial, it is always at zero for the particle-hole symmetric (PHS) model. The above analytical results for $|\omega_{nQD}^0|$ nicely agree with recent experimental data on two~\cite{Dvir2023} and three site~\cite{Bordin2025} Kitaev chains in a different geometry. These results are also consistent with recent theoretical observations of topology appearing in chains of increasing length~\cite{Dourado2026}.

\section{Majorana Representation: finite length chain}\label{sec:majoranarepr}

In this Section, we introduce the Majorana representation of the Hamiltonian $H_{2n\Psi}$. This representation is a convenient tool for analysing the topological features of the model. We construct Majorana operators out of electron creation and annihilation operators at each site $j$ and for each sublattice $\alpha=A,B$. Four  types of Majorana operators are defined as

\begin{align}\label{maj-gam}
\gamma_{\alpha,j}&=\frac{1}{\sqrt{2}} \left(d_{\alpha,j} + d_{\alpha,j}^\dagger\right),
\\
\eta_{\alpha,j}&=\frac{-i}{\sqrt{2}} \left(
d_{\alpha,j} - d_{\alpha,j}^\dagger
\right),
\label{maj-eta}
\end{align}
where $\alpha=A,B$ denotes the sublattice and $j$ counts dimmers. The Hamiltonian (\ref{hN2}) can be rewritten as
\begin{align}
H^M_{2n\Psi} = \frac{i}{2}&\sum_{j} \Big[
m_A\,\gamma_{A,j}\eta_{A,j}
+m_B\,\gamma_{B,j}\eta_{B,j}
\nonumber \\
&-2t_{+}\,\eta_{A,j}\gamma_{B,j}
+2t_{-}\,\gamma_{A,j}\eta_{B,j}
\nonumber \\
&+2t_{+}\,\eta_{B,j}\gamma_{A,j+1}
-2t_{-}\,\gamma_{B,j}\eta_{A,j+1}
\Big].
\label{MH}
\end{align}
Recall that $m_A = \mu + V_z$, $m_B = \mu - V_z$, and we have defined the effective hoppings  $t_{\pm} = t_{so} \pm \Delta_{CAR}$. In the adopted representation, both on-site terms and the effective hoppings show a staggered-like structure in the combined dimer $j$, sublattice, Majorana type $(\eta_{\alpha,j}; \gamma_{\beta,j})$ space.
The general properties of the long chain in this representation are analysed in \ref{Ntopology}, while in the next Section we specialise to the shortest chain exhibiting the mentioned structure, namely a four-QD chain which consists of two dimers.

\section{Chiral Majorana representation for two dimmers}

For a short chain with 4QDs, the dimer summation index $j$ in (\ref{hN2}) takes on only two values $j=0,1$. Thus, in fact, we are dealing with a two-dimer system. This is the shortest chain, which shows a two-sublattice structure and allows for a periodic chiral Majorana representation.

\begin{figure}
\centerline{\includegraphics[width=1\linewidth,clip]{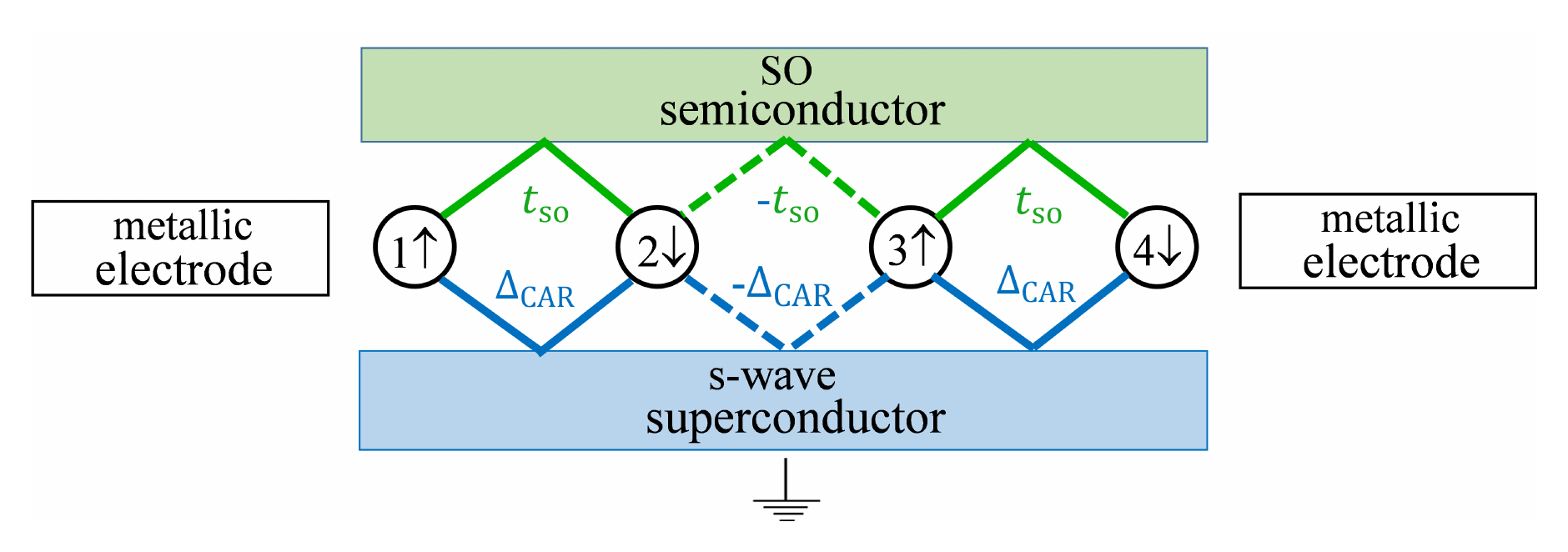}}
\caption{Scheme of the 4QD chain with a staggered spin configuration and alternating inter-dot couplings. The notation in the picture corresponds to the $\Psi$ state. For the $\Phi$ state, the spins at  all quantum dots would be reversed.}
\label{scheme-dimer}
\end{figure}

There exist many possible choices of Majorana bases to represent the Hamiltonian $H^M_{2n\Psi}$, and each of them displays different symmetry. We have found that the chiral representation is a convenient one and suits our purposes here.
This representation is given by the   $\Gamma_\gamma = (\gamma_{A,0}, \gamma_{B,0}, \gamma_{A,1}, \gamma_{B,1})^T$
and  $\Gamma_\eta = (\eta_{A,0}, \eta_{B,0}, \eta_{A,1}, \eta_{B,1})^T$. Note that we first group the consecutive  $\gamma$-type Majoranas and later those of $\eta$-type.
Combining these two sets of states, we get a full 8-component Majorana spinor for a whole chain
\begin{align}
\Psi_{c} &= (\Gamma_\gamma; \Gamma_\eta)^T .\label{chib}
\end{align}
In this chiral basis, the Hamiltonian (\ref{MH})  is written as $H^M_{4\Psi} = \frac{i}{2} \Psi_{c}^T \mathcal{A}_{c} \Psi_{c}$, where the matrix $\mathcal{A}_{c}$ takes a purely block off-diagonal form
\begin{equation}
\mathcal{A}_{c} =
\begin{pmatrix}
0 & h_{4} \\
-h_{4}^T & 0
\end{pmatrix},
\label{chim}
\end{equation}
where $h_{4}$ is a $4 \times 4$ real matrix containing all the coupling parameters. It reads
\begin{equation}
h_{4} =
\begin{pmatrix}
m_A & t_- & 0 & 0 \\
t_+ & m_B & -t_- & 0 \\
0 & -t_+ & m_A & t_- \\
0 & 0 & t_+ & m_B
\end{pmatrix}.\label{hmatrix}
\end{equation}

Finally, the matrix Green function for a chain tunnel coupled to two external normal-metallic electrodes ($c.f.$ Fig.~\ref{scheme-dimer}) can be written as
\begin{eqnarray}
   [G^r_c(\omega)]^{-1}=\omega \mathbf{1}_{8\times8}-\frac{i}{2}\mathcal{A}_c-\Sigma(\omega),
   \label{majGF}
\end{eqnarray}
where we now consider a chain in contact with the external electrodes. This is taken into account by adding the self-energy
\begin{equation}
\Sigma(\omega)=-i \operatorname{diag}(\gamma_L,0,0,\gamma_R,\gamma_L,0,0,\gamma_R)\equiv -i~\Gamma
\end{equation}
which is calculated in the wide-band approximation. It is important to note that the various elements of the Green function (\ref{majGF}) describe different transfer processes along the chain.

We note in passing that another useful representation, called the SSH representation due to the analogy with the corresponding Su-Schrieffer-Hegger model~\cite{Su1979}  would be represented by the spinors $\Gamma_+ = (\eta_{A,0}, \gamma_{B,0}, \eta_{A,1}, \gamma_{B,1})^T$ and $\Gamma_- = (\gamma_{A,0}, \eta_{B,0}, \gamma_{A,1}, \eta_{B,1})^T$. In this representation $\Psi_{SSH}=(\Gamma_+;\Gamma_-)^T$, the Hamiltonian also has a block form, however with block-diagonal and off-diagonal terms. The diagonal matrices depend on $t_{\pm}$ parameters, while the off-diagonal matrices have a structure like $\operatorname{diag}(-m_A,m_B,-m_A,m_B)$. This representation is more convenient if one wants to study $e.g$  a role of the mass terms $m_A$ and $m_B$.

\begin{figure}
\centerline{\includegraphics[width=1\linewidth,clip]{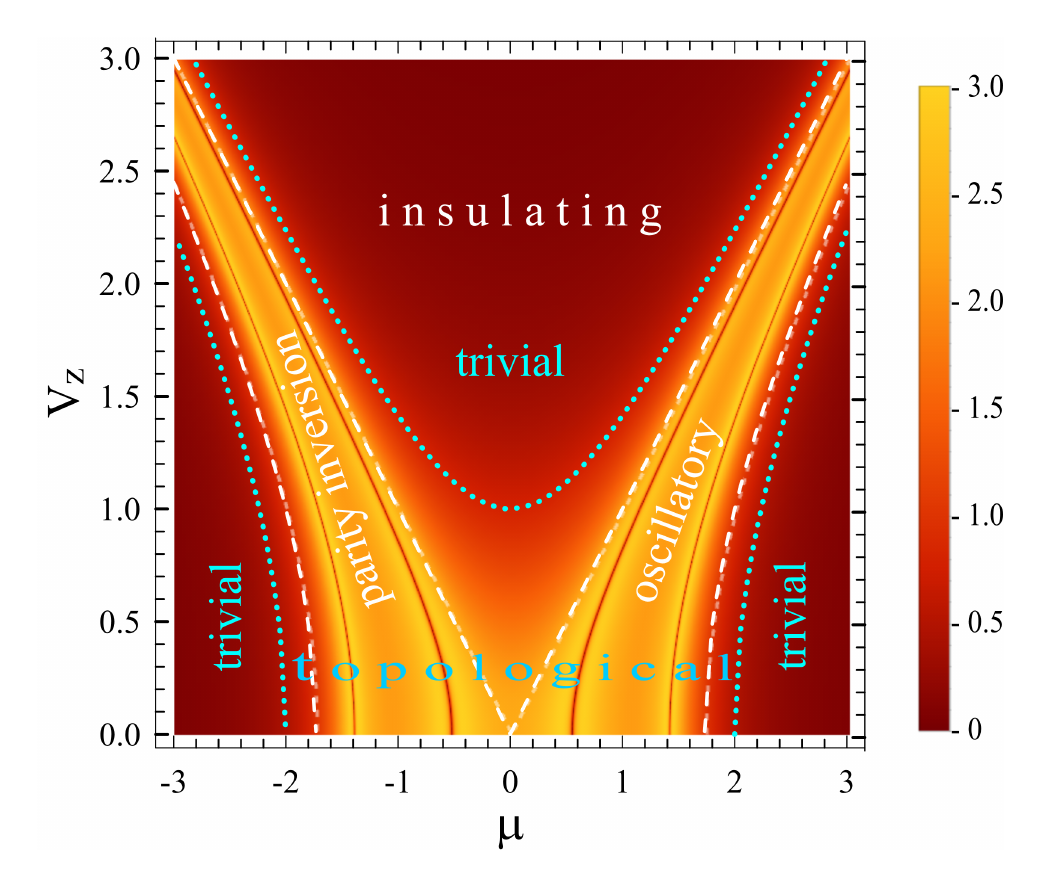}}
\caption{
Phase diagram of the open 4QD Majorana chain in the
$(\mu,V_z)$ plane determined from the nonlocal zero-energy chiral Majorana
Green function $\log\!\left(1+|G_{\gamma_{A0}\eta_{B1}}^r(0)|\right)$
describing the transfer $\gamma_{A1} \rightarrow \eta_{B1} \rightarrow
\gamma_{A2} \rightarrow \eta_{B2}$,  for $t_{so}=1$, $\Delta_{CAR}=0.5$ and $\gamma_L=\gamma_R=0.05$.
 The colour scale represents the strength of coherent nonlocal Majorana
propagation between opposite ends of the chain. Bright regions indicate
strong edge-to-edge coherence, enhanced teleportation-like transport,
and strongly hybridised finite-size Majorana states, while dark regions
correspond to suppressed nonlocal propagation and localised edge-mode
behavior.
In the two regions symmetric with respect to $\mu=0$  located between $V_z=|\mu|$ and $V_z^2=\mu^2-4(t_{so}^2-\Delta_{CAR}^2)$, Eq.(\ref{eq:osc}),  (denoted by white dashed lines), the oscillatory standing wave solutions exist. Inside the oscillatory sector, parity-crossing occurs where the nonlocal Green function
resonantly changes sign (bright-dark arcs).
The cyan dotted curves present the topological phase boundaries: $V_{z,0}^2 = \mu^2 + 4\Delta_{CAR}^2$ and $V_{z,\pi/2a}^2 = \mu^2 - 4t_{so}^2 $, Eqs.\eqref{vz0}-\eqref{vzpi}, determined by means of the topological invariant $\nu$ (with help of the Pfaffians) for the infinite chain.}
\label{phasediagGreen}
\end{figure}

\section{Majorana Green functions: signatures of topology for a chain with 4QDs}\label{sec:topo-phase}

Our main interest is in the spatial structure of zero-energy modes and the signatures of topology in short chains; thus, we consider only zero-energy Green functions. It is the  $G_{\gamma_{A0}\eta_{B1}}^r(\omega=0)$ element of the Majorana Green function, which is of main interest. It describes the coherent propagation across the whole chain, $\gamma_{A1}\rightarrow \eta_{B1} \rightarrow \gamma_{A2} \rightarrow \eta_{B2}$.   In Fig.~\ref{phasediagGreen} we plot this Green function calculated for $\omega=0$ on the ($\mu,V_z$) plane. In fact, to avoid singularities, we are plotting the function
$\log\!\left(1+|G_{\gamma_{A0}\eta_{B1}}^r(0)|\right)$ for a chain with $t_{so}=1$, $\Delta_{CAR}=0.5$ and $\gamma_L=\gamma_R=0.05$.

The ($\mu,V_z$) map in Fig.~\ref{phasediagGreen} acts as a direct visualisation of the efficiency of Majorana propagation between the opposite ends of the 4QD chain. This map exhibits a rich structure composed of several distinct regions separated by sharp boundaries. While these boundaries emerge naturally from the Green-function analysis, their physical origin is not immediately obvious from the propagator itself. To understand the mechanisms responsible for the observed features, we analyse the map as a topological phase diagram using several complementary approaches, all of which consistently describe the crossover from a strongly hybridised finite-size Majorana regime to the formation of localised Majorana edge states. The various criteria arise from the transfer-matrix approach, Green-function analysis, parity-crossing conditions, and level inversion, providing a unified picture of the emergence of topology in finite systems \cite{Alicea2012, Beenakker2013}.

Remarkably, all these methods identify the same characteristic
boundaries and therefore provide a unified interpretation of the
structures visible on the Green-function map. Together, they show how the finite 4QD chain evolves between regimes of oscillatory
Majorana states, strongly hybridised finite-size modes, and
localised edge Majorana states.

In Fig.~\ref{phasediagGreen}, one can clearly distinguish two different regions: the upper and the left/right triangles, corresponding to two different spectral dependencies with insulating and oscillating zero-energy states exhibited in Fig.~\ref{spectrum}.

\paragraph{Oscillatory regime}

The transfer matrix governs the fundamental spatial structure of the zero-energy solutions, as its eigenvalues can distinguish between oscillatory and localised wavefunctions \cite{DeGottardi2013,DeGottardi2013PRL}.
This happens when the eigenvalues of the transfer matrix $T$ in Eq. (\ref{T}) change from real to imaginary values (see  \ref{app:transfer}). From this, we extract the exact boundaries
\begin{equation}
m_Am_B=0,
\qquad
m_Am_B=4t_+t_-,
\label{eq:osc}
\end{equation}
which separate regions with real transfer-matrix roots from those with complex roots. These boundaries are presented as the white-dashed curves in Fig.~\ref{phasediagGreen}.
Inside this bounded region, the transfer-matrix eigenvalues form a complex-conjugate pair, $\lambda_\pm = |\lambda|e^{\pm iq}$,
and the zero-energy modes acquire an oscillatory character. For a finite open chain, these oscillatory transfer-matrix modes interfere coherently, producing standing-wave Majorana states. Outside this region, the transfer-matrix eigenvalues become purely real ($\lambda_\pm \in \mathbb{R}$), resulting in evanescent solutions and exponentially localized spatial envelopes.
Crucially, while the bulk TRIM Pfaffians strictly assume the limit ($n \to \infty$), the transfer matrix approach remains valid for any finite system size $n$, taking into account Majorana wavefunction overlap and finite-size splitting as well.

\paragraph{Parity crossing}

The concept of ``parity crossing'' was originally formulated by Kitaev \cite{Kitaev2001} to describe the emergence of Majorana-bound states.
Because these Majorana states reside at the Fermi energy (zero energy), populating or depopulating them requires zero thermodynamic work. Consequently, the many-body ground state of the system becomes doubly degenerate, consisting of one state with an even number of fermions and another with an odd number of fermions. As one tunes the Hamiltonian parameters (such as the chemical potential or magnetic field) across the topological phase boundary, the lowest-energy single-particle state crosses through zero energy. At this critical point, the many-body ground state swaps its fermion parity. Mathematically, the existence of this zero-energy state is given by the vanishing of the determinant of the single-particle Hamiltonian \cite{Prada2012}.
While the strict topological phase transition occurs only in the $n \rightarrow \infty$  limit, finite-size precursors of this physics are distinctly visible in short systems.

For the finite 4QD chain, the parity-crossing condition
is encoded in the chiral block $h_{4}$, Eq.~(\ref{hmatrix}), given explicitly by the determinant
\begin{equation}
\det(h_{4}) =
(m_Am_B)^2
-3m_Am_B\,t_+t_-
+(t_+t_-)^2.
\end{equation}
The exact zero-energy parity crossings occur when this determinant vanishes, yielding the golden-ratio critical points:
\begin{equation}
m_Am_B
=
\frac{3\pm\sqrt5}{2}\,t_+t_-.
\end{equation}

The physical significance of these conditions can be seen in nonlocal transport propagators. The zero-energy Green function is obtained by inverting the system's Hamiltonian; in the chiral basis, the nonlocal transport components are proportional to the inverse of the chiral block, $G^r \propto h_{4}^{-1} = \text{adj}(h_{4}) / \det(h_{4})$. Consequently, the condition $\det(h_{4}) = 0$ corresponds precisely to the energy levels in the isolated system (presented in Fig.~\ref{spectrum}). Near the parity
crossings the poles approach zero energy, producing a strong enhancement
of the nonlocal propagators and of the associated crossed-Andreev and
elastic cotunneling amplitudes
\cite{Prada2012,Rainis2013,Prada2017}.

This pole structure of the Green function dictates the primary visual features of the transport phase diagram. In Fig. (\ref{phasediagGreen}) we plotted quantity, $\log\left(1+|G_{\gamma_{A1}\eta_{B2}}^r(0)|\right)$, specifically chosen to capture the divergences. Without dissipation, the amplitude $|G_{\gamma_{A1}\eta_{B2}}(0)|$ would diverge
at the determinant zeros, $\det(h_{4}) = 0$. The logarithmic scaling compresses this massive dynamic range, translating the mathematical poles into sharp, bright-dark continuous arcs in the intensity profile. These striking, bright-dark features emerge precisely because the Green function fundamentally changes its character during the parity crossover, leaving a stark visual signature within the oscillatory sector of the phase diagram.

Furthermore, the phase diagram captures the physically realistic dissipative topology. The finite electrode couplings ($\gamma_L$ and $\gamma_R$) introduce a self-energy that shifts the true poles slightly off the real axis into the complex energy plane. As a result, the exact mathematical divergence of the isolated Green function is regularised. The determinant zeros no longer cause a strict singularity; instead, they produce a finite, broadened resonant peak. The bright-dark teleportation bands observed in the plot are exactly these regularised poles. As the dissipation $\gamma$ increases, these arcs widen, directly visualising how the leads smear the discrete finite-size spectrum and broaden the transmission resonances.

\section{Conclusion}\label{sec:concl}

We have considered a finite chain of $n$ quantum dots proximitised to a standard s-wave, BCS-like superconductor and a semiconducting system with strong spin-orbit coupling. Our geometry differs from the one already realised experimentally~\cite{Dvir2023,Bordin2024,Bordin2025} in which two, respectively three quantum dots are coupled {\it via} superconductors. We consider a sandwich geometry in which a whole chain of QDs is proximitised to both a superconductor and a semiconductor with strong spin-orbit coupling. Allowing for crossed Andreev reflections and spin-flipping scattering involving only neighbouring quantum dots, we have analysed the appearance and properties of Majorana zero modes.
The spectrum of short chains uniquely indicates increased protection in longer chains. The deviation of parameters from the sweet spot $t_{so}=\Delta_{CAR}$ leads to slow increase of energies as  $|\omega_{nQD}^0|\propto\; (t_{so}-\Delta_{CAR})^{n/2}/(2t_{so})^{n/2-1}$ for even number $n$.

We have also studied the spacial character of MZM, as discussed in details in \ref{app:transfer} and identified two different regions in parameter space. In one of them the Majorana wave function exhibits damped spatial oscillations. This region is placed between the boundaries given by \eqref{eq:osc}. Since we consider the case $\Delta_{CAR}^2 < t_{so}^2$ ,
the oscillatory region is entirely contained within the topological phase
\begin{align}
0 < m_Am_B <4&\left(t_{so}^2-\Delta_{CAR}^2\right)
\subset \nonumber\\
& -4\Delta_{CAR}^2 < m_Am_B <4t_{so}^2.
\end{align}
The other portion of the phase diagram, in which long chain is topological, supports non-oscillatory, exponentially localised zero modes at chain ends.

We have shown  that all these features can be spotted in short chains by studying retarded Green function in the chiral Majorana representation. However, one has to look at the appropriate function, namely that one, which probes the whole chain (see  Section \ref{sec:topo-phase}). The color coded plot of the Green function shown in Fig. \ref{phasediagGreen}, nicely agrees with the expected exact phase diagram of the long chain.
In particular the strong features in the Green function uniquely describe the parity crossing and  the regions of oscillatory zero modes. One has to note that the coupling of the external quantum dots to  reservoirs introduces damping and results in smearing of some features on the phase diagram. However, this is in agreement with experiments, which also introduce some dissipation. In summary,
the retarded Green function in chiral Majorana representation is a powerful tool to analyse properties of MZM in short chains.

\appendix

\section{Two Hilbert spaces}\label{sec:twoHilbert}

To proceed, we consider a chain consisting of an even number of quantum dots ($2n$) and decompose the Hilbert space into two independent subspaces.
In analogy to the procedure described in Ref. \cite{Bulka2026} we separate the chain into two staggered sections represented by the Nambu spinors
$$\Psi^\dagger=(c^\dagger_{1\uparrow},c^\dagger_{2\downarrow},c^\dagger_{3\uparrow},c^\dagger_{4\downarrow}, \dots)$$and
$$\Phi^\dagger=(c^\dagger_{1\downarrow},c^\dagger_{2\uparrow},c^\dagger_{3\downarrow},c^\dagger_{4\uparrow}, \dots).$$
As a result the Hamiltonian (\ref{ham:prox}) in this basis can be rewritten as
\begin{align}
 H_{prox}=\frac12 &\left( \Psi^\dagger \mathcal{H}_{\Psi\Psi}\Psi +\Phi^\dagger \mathcal{H}_{\Phi\Phi}\Phi \right. \nonumber\\
&\left.+\Psi^\dagger \mathcal{H}_{\Psi\Phi}\Phi +\Phi^\dagger \mathcal{H}_{\Phi\Psi}\Psi \right).
\label{Hproxupdown}
\end{align}

Close inspection of the mixing Hamiltonian matrix  $H_{\Psi\Phi}$ shows that it contains only two parameters. The first one is the direct hopping $t$ and the second is the induced on-site pairing $\Delta_{LAR}$. Assuming the absence of them both, allows us to neglect the mixing altogether. The resulting Hamiltonian is a sum of two terms defined in respective Hilbert spaces $\Psi$ and $\Phi$. With this assumption, one ends up with the separation of two subspaces. The Hamiltonian is written as a sum of two terms corresponding to two independent subspaces, to be denoted simply $H_{2n\Psi}$ and $H_{2n\Phi}$, and it is enough to analyse only a single subspace.

\section{Real-Space Transfer Matrix Approach}\label{app:transfer}

To reveal a real-space structure of the Majorana wave functions, one uses an approach based on the transfer matrix technique.
For the studied Majorana chain in a chiral Majorana representation $\Psi_{c} = (\Gamma_\gamma; \Gamma_\eta)^T$, Eq.\eqref{chib},
we evaluate the eigenvalue equation $\mathcal{H}_{4\Psi}^M\Psi_c = E \Psi_c$,
which at zero energy gives two decoupled equations $h_{4} \Gamma_\eta = 0$ and $h_{4}^T \Gamma_\gamma = 0$. Confining attention to the $\eta$ sector, we have  the following set of equations
\begin{align}
    m_A\eta_{Aj}+t_{-}\eta_{Bj}-t_{+}\eta_{Bj-1}=0,\\
    m_B\eta_{Bj}+t_{+}\eta_{Aj}-t_{-}\eta_{Aj+1}=0.
\end{align}
Elimination of Majorana operators belonging to the $B$ sublattice results in $\Phi_{j+1}=T\Phi_j$, where the transfer matrix
\begin{equation}
    T=
    \begin{pmatrix}
\frac{2t_{-}t_{+}-m_Am_B}{t_{-}^2} & -\frac{t_{+}^2}{t_{-}^2} \\
1& 0
\end{pmatrix}.
\label{T}
\end{equation}
From this equation, we read off the transfer matrix relating the wave function between two neighbouring unit cells. The transfer matrix relating more distant site: $e.g.$ the last unit cell to the first one would be a product of all intermediate transfer matrices. However, with parameters independent of a site or dimer number, the information about the spatial structure of arbitrarily distant unit cells along the chain is encoded in $T$. Using the periodicity of the structure, we relate the properties of the transfer matrix to the spatial character of Majorana wave functions.

The eigenvalues of the transfer matrix T satisfy the characteristic equation
\begin{equation}
\lambda^2-a\lambda+b=0,
\end{equation}
where $a=(2t_+t_- - m_A m_B)/t_-^2$, $b=t_+^2/t_-^2$. Its solutions are $\lambda_\pm = (a\pm\sqrt{\Delta})/2$, with the discriminant
\begin{equation}
\Delta =\frac{ m_A m_B \left(m_A m_B-4t_+t_-\right)}{t_-^4}.\label{discriminant}
\end{equation}
The sign of $\Delta$ determines the spatial character of wave-functions, and $\Delta=0$ gives the boundary \eqref{eq:osc}.

The bulk criteria can be precisely related to a real-space transfer matrix
$T$ governing the recursive spatial envelope of the zero-energy modes
($\Phi_{j+1}=T\Phi_j$). A bulk topological phase transition occurs when
one eigenvalue $\lambda$ of $T$ crosses the unit circle
($|\lambda|=1$), separating exponentially decaying and growing
zero-energy solutions. This unit-circle crossing is equivalent to the closing of the bulk excitation gap at a time-reversal-invariant momentum (TRIM)
and to a change of the $\mathbb{Z}_2$ topological invariant
\cite{Akhmerov2011,Brouwer2011,Tewari2012}.

Using Bloch's theorem, together with analytical continuation, the transfer matrix eigenvalues can be directly identified with
the generalised Bloch factor: $\lambda \equiv z = e^{i 2ak}$,
which describes the complex phase accumulated across a unit cell. For extended bulk states, we have $|z|=1$, whereas $|z|\neq 1$ corresponds to exponentially growing or decaying solutions of spatially localised states.
This gives a rigorous correspondence between spatial localisation in the transfer-matrix formalism and the geometric topology of the bulk bands. In particular, the crossing of an eigenvalue through the unit circle at a TRIM shows a change in the topological invariant and marks the onset of a topological phase transition.

\section{Momentum-Space Majorana Hamiltonian and Its Topology}
\label{Ntopology}
Our aim here is to apply standard techniques~\cite{Sau2012} and examine the topology of the long chain with $n\rightarrow\infty$.
In this limit, we transform the Majorana Hamiltonian $H^M_{2n\Psi}$, Eq.(\ref{MH}), applying the canonical Bloch representation~\cite{Cayssol2021} of the real-space Majorana operators (where positions of sites are appropriately shifted with $x_{A,j} = 2ja$ and $x_{B,j} = 2ja+a$, $a$ denotes an inter-dot distance).
The momentum-space Hamiltonian in Majorana representation reads
\begin{equation}
H_M = \frac{i}{4} \sum_k \Gamma_k^\dagger \mathcal{H}_M(k) \Gamma_k,
\end{equation}where $\Gamma_k = (\gamma_{A,k}, \eta_{A,k}, \gamma_{B,k}, \eta_{B,k})^T$.
 The exact $4 \times 4$ skew-Hermitian Hamiltonian matrix takes the block
\begin{equation}
\mathcal{H}_M(k) =
\begin{pmatrix}
m_A i\sigma_y & Q(k) \\
-Q^\dagger(k) & m_B i\sigma_y
\end{pmatrix},
\label{hmm}
\end{equation}
where the off-diagonal inter-sublattice coupling matrix is given by:
\begin{equation}
Q(k) =
\begin{pmatrix}
0 & 2t_- e^{ika} - 2t_+ e^{-ika} \\
-2t_+ e^{ika} + 2t_- e^{-ika} & 0
\end{pmatrix}.
\end{equation}

\subsection{Bulk Topological Phase Criterion}\label{app:bulktopo}

One way to define the topologically non-trivial phase is to calculate~\cite{Kitaev2001} the topological invariant $\nu$, sometimes called the Majorana number $\mathcal M$. For the above system, it is defined as~\cite{Kitaev2001} \begin{equation}
    \nu=\mathcal{M}=\rm{sgn}[\rm{Pf}(\mathcal{H}_M(0))\rm{Pf}(\mathcal{H}_M(\pi/2a))],
\end{equation}
where $\rm{Pf(A)}$ denotes the Pfaffian of matrix A. We used $k=\pi/(2a)$ because the Brillouin zone has been halved due to period doubling in real space. Denoting $q_1=2t_- e^{ika} - 2t_+ e^{-ika}$ and $q_2=-2t_+ e^{ika} + 2t_- e^{-ika}$ we calculate the Pfaffian of the matrix  in Eq.~(\ref{hmm}) as $\rm{Pf}(\mathcal{H}_M(k))=m_A m_B + q_1 q_2$. Evaluating the Hamiltonian matrix for $k=0$ and $k=\pi/(2a)$ we get the topological index
\begin{equation}
    \nu=\rm{sgn}[(m_A m_B + (t_- - t_+)^2)(m_A m_B-(t_- + t_+)^2)].
\label{eq;toponu}
\end{equation}
The value $\nu=-1$ signals the topologically non-trivial phase, whereas in the trivial phase $\nu=+1$.

Evaluating the Pfaffian of $\mathcal{H}_M(k)$ at these points yields the critical Zeeman field boundaries:
\begin{align}
V_{z, 0}^2 &= \mu^2 + 4\Delta_{CAR}^2  \;\; \text{-  at } k=0,\label{vz0}\\
 V_{z,\pi/2a}^2 &= \mu^2 - 4t_{so}^2  \quad\quad \text{- at } k=\pi/2a\label{vzpi}.
\end{align}
The topologically non-trivial phase (Kitaev invariant $\nu = -1$)~\cite{Kitaev2001} occurs when the two mass gaps have opposite signs. This imposes the  boundaries for the topological phase:
\begin{equation}
 \mu^2 - 4t_{so}^2 < V_z^2 < \mu^2 + 4\Delta_{CAR}^2.
\end{equation}
This condition shows that a sufficiently strong Zeeman field $V_z$ destroys the topological phase, closing the gap at the second TRIM. The boundaries of the topological region are shown in Fig. (\ref{phasediagGreen}) by cyan dotted curves.

\subsection{Summary}

Within the transfer matrix approach, the spatial character of the Majorana edge states is determined by the nature of the discriminant $\Delta$, Eq. \eqref{discriminant}, and the wave-functions show oscillatory behaviour for $\Delta<0$, which gives the boundaries \eqref{eq:osc}. Since we consider the case $\Delta_{CAR}^2 < t_{so}^2$, the oscillatory region is entirely contained within the topological phase
\begin{align}
0 < m_Am_B <4&\left(t_{so}^2-\Delta_{CAR}^2\right)
\subset \nonumber\\
& -4\Delta_{CAR}^2 < m_Am_B <4t_{so}^2.
\end{align}
Therefore, the same bulk characteristic polynomial simultaneously encodes two complementary aspects of the system. The values of $\det \mathcal{H}_M(k)$ at the two TRIM determine the Majorana topological invariant and the boundaries of the topological phase, while the discriminant determines whether the edge states decay monotonically or exhibit damped spatial oscillations. Consequently, every oscillatory Majorana state is topological, whereas a finite portion of the topological phase supports purely exponentially localised Majorana modes without oscillations. According to the theory~\cite{Souto2023}, the oscillatory and localised behaviour can be distinguished experimentally.

\acknowledgments

 We would like to thank prof. T. Doma\'nski for discussions. This work has been partially supported by the National Science Centre, Poland (``Weave'' programme) through grant no. 2022/04/Y/ST3/00061.

\bibliography{splitterSO}

\end{document}